\documentclass[12pt,preprint]{aastex}
\newcommand{\Msun}{\rm M_\odot}
\newcommand{\Rsun}{\rm R_\odot}

\begin{document}

\title{The Formation of Cataclysmic Variables with Brown Dwarf Secondaries}

\author{Michael Politano}

\affil{Department of Physics, Marquette University, P.O. Box 1881, Milwaukee, WI 53201-1881}

\begin{abstract}
The present-day formation of cataclysmic variables (CVs) with brown dwarf (BD) secondaries (0.013 $\Msun \le$ M$_s \le$ 0.075 $\Msun$) is investigated using a population synthesis technique.  Results from the latest, detailed models for BDs have been incorporated into the population synthesis code.  The present-day orbital period distribution of zero-age CVs (ZACVs) that form with BD secondaries is calculated.  For our models, we find that ZACVs with BD secondaries have orbital periods in the range 46 min to 2.5 hrs.  We also find that ZACVs with BD secondaries comprise 18\% of the total, present-day ZACV population.  In addition, we find that 80\% of ZACVs with BD secondaries have orbital periods {\it less than} 78 minutes.  This implies that 15\% of the present-day ZACV population should have orbital periods {\it shorter} than the observed orbital period minimum for CVs.  We also investigate the dependence of the present-day formation rate of CVs with BD secondaries on the assumed value of the common envelope efficiency parameter, $\alpha_{CE}$, for three different assumed mass ratio distributions in ZAMS binaries.  Surprisingly, we find that the common envelope process must be extremely inefficient ($\alpha_{CE} <$ 0.1) in order for CVs with BD secondaries not to be formed.  Finally, we find that the progenitor binaries of ZACVs with BD secondaries have ZAMS orbital separations $<$ 3 AU and ZAMS primary masses between $\sim$ 1-10 $\Msun$, with $\sim$ 75\% of the primary masses less than $\sim$ 1.6 $\Msun$.  Interestingly, these ranges in orbital separation and primary mass place the majority of the progenitor binaries {\it within the so-called ``brown dwarf desert.''}  The implications of these results are discussed in the context of V485 Cen and 1RXS J232953.9+062814, the only known CVs with orbital periods shorter than the period minimum, CVs above the period minimum that are strongly suspected of containing BD secondaries (such as WZ Sge), common envelope evolution involving very low mass secondaries, and the BD mass function in ZAMS binaries.
\end{abstract}

\keywords{binaries: close---novae, cataclysmic variables---stars: evolution---stars: formation---stars: fundamental parameters---stars: low-mass, brown dwarfs---stars: luminosity function, mass function}

\section{Introduction}

The formation of cataclysmic variables (CVs) has been studied in detail by several authors \citep{pol88,pol96,dek92,han95,how01}.  However, there has been only one detailed study of the formation of CVs with brown dwarf (BD) secondaries to date, that of \citet{pol88,pol96}.  This study used (by now) outdated physics for the radius-mass relationship of the BD and assumed that the primordial distributions of main-sequence (MS) + BD binaries was the same as the primordial distribution of MS + MS binaries.  A more recent study considered the secular evolution of CVs with BD secondaries using up-to-date models for the brown dwarfs \citep{kol99}.  However, this study did not calculate the formation of CVs with brown dwarf secondaries, but rather calculated the CV birthrate as a function of secondary mass for MS stars and extrapolated it into the BD regime.
 
Significant advances both in our theoretical and observational understanding of brown dwarfs have occurred in the past five years \citep[e.g.,][]{bur97,cb97,cb00,kir00,zap00,bej01,bur02,dah02}. Such important advances now offer an excellent opportunity to apply these models to the low mass secondaries in CVs.  We have incorporated into our population synthesis code \citep{pol96} current models of low-mass stars and BDs from the Lyon group.  Their evolutionary code includes improved internal physics \citep{cb97,cb00} and outer boundary conditions based on non-gray atmosphere models \citep[see][]{all97}, appropriate to describe the complex structure of low mass stars and brown dwarfs.  These improvements in the present theory have led to a better understanding of the properties of M-dwarfs of solar metallicity \citep{bar98} and yield excellent agreement with globular cluster lower main sequences \citep{bar97}.  The models extended to the brown dwarf regime \citep{cbet00,bar03} successfully reproduce the near-IR photometric and spectroscopic properties of the cool brown dwarf GL 229B \citep{all97}, as well as the young and hot brown dwarfs observed recently in young clusters \citep[cf.][]{zap97,zap00,mart98,bej01}.  

Further motivation for the present work is provided by the existence of two unusual systems:  V485 Cen, a CV with an orbital period of 59 min \citep{aug93,aug96,ole97}, and 1RXS J232953.9+062814, a CV with an orbital period of 64 min (hereafter RX2329+06; \citealp{jin98,uem01,uem02,tho02}).  In the standard model of CV secular evolution, a CV evolves to shorter orbital periods, reaches a minimum period, and then begins to evolve back toward longer orbital periods (\citealp{pac81,rap82}; see also \citealp{kin88} for a review).  The orbital periods in both V485 Cen and RX2329+06 are shorter than either the observed (78 min) or theoretically-predicted ($\sim$ 70 min) orbital period minimum \citep[cf.][]{kol99} for CVs that contain relatively unevolved main sequence secondary stars, and thus cannot be explained within the standard model of CV secular evolution.  

There are objects related to CVs with orbital periods less than 78 min, namely the AM CVn stars (P$_{\rm orb}$ $\sim$ 20-50 min; \citealp{rit03,nel01}).  AM CVn stars are believed to contain a white dwarf and a degenerate helium white dwarf or a semi-degenerate helium star (see \citealp{ull94} for a review; also \citealp{nel01} and references therein).  This model is consistent with the notable very weak hydrogen lines in the spectra of these stars.  However, both V485 Cen and RX2329+06 have orbital periods longer than known AM CVn stars, and both systems contain hydrogen lines in their spectra \citep{aug93,aug96,tho02}.  The presence of hydrogen lines in the spectra of V485 Cen and RX2329+06 thus precludes a model for these systems in which the secondary is a He star.

It has been suggested instead that both V485 Cen and RX2329+06 contain a secondary in which a significant fraction of hydrogen in the core had been processed into helium {\it prior} to mass transfer \citep{aug96,sch02,tho02,uem02}.  By the time such a system reaches short orbital periods, significant erosion of the secondary will have exposed the He-rich core.  Calculations of the secular evolution of such CVs predict that these systems can reach orbital periods as short as those observed in V485 Cen and RX2329+06 \citep{sch02, tho02}.  However, \citet{pol88,pol96} also found that CVs that {\it form} with substellar secondaries could have orbital periods, at birth, as short as 34 min. Since his calculations relied on outdated physics for the substellar secondaries, his result needs to be confirmed using population synthesis calculations that contain up-to date models for the substellar secondaries in order to determine whether a CV that formed with a BD secondary is also a possible model for V485 Cen and RX2329+06.

We also examine in this paper the dependence of the formation rate of CVs with BD secondaries on the assumed value of the common envelope efficiency parameter, $\alpha_{CE}$.  In the absence of readily incorporable results from detailed hydrodynamical models of the common envelope (CE) phase (cf. \citealp{ibe93,taa00} for recent reviews), population synthesis calculations have relied on a simple energy argument to relate the pre-CE and post-CE orbital parameters.  Essentially, this argument states that when some fraction of the orbital energy released in the spiral-in process equals the initial binding energy of the primary's envelope, the CE becomes unbound and is presumably (somehow) ejected \citep{tut79}.  Although the exact prescription varies from author to author, a typical expression used in population synthesis calculations is given below \citep[cf.][]{ibe85,dis94,pol96}: 

\begin{equation}
  -\alpha_{CE}(\frac{GM_cM_s}{2a_f} - \frac{GM_pM_s}{2a_i})
     = -\frac{\gamma_1GM_eM_e}{R_i} - \frac{\gamma_2GM_eM_c}{R_i}\,
\end{equation}
\linebreak
where M$_{\rm p}$ is the total primary mass at the onset of the CE phase, M$_{\rm s}$ is the mass of the secondary, M$_{\rm c}$ is the core mass of the primary, M$_{\rm e}$ is the envelope mass of the primary (M$_{\rm e}$ = M$_{\rm p}$-M$_{\rm c}$), R$_{\rm i}$ is the radius of the primary at the onset of the CE phase (and is therefore equal to the radius of the primary's Roche lobe), a$_{\rm i}$ and a$_{\rm f}$ are the pre-and post-CE orbital separations, respectively, $\gamma_1$ and $\gamma_2$ are dimensionless factors related to the structure of the giant star, and $\alpha_{CE}$ is a parameter that measures the efficiency at which orbital energy is able to be used in unbinding the CE.

From equation 1, we see that the orbital energy depends linearly on the mass of the secondary.  Because of the small secondary mass, in considering the spiral-in of a BD secondary into the giant envelope, there may simply not be enough of an energy reservoir available in the orbit to unbind the envelope if the transfer of energy from the orbit is too inefficient.  Instead the CE phase may result in merger of the component stars.  While \citet{pol96} found that CVs with BD secondaries were able to be formed, he assumed that the transfer of energy was 100\% efficient (i.e., $\alpha_{CE}$ = 1).  This has been a fairly standard assumption in population synthesis studies of CVs \citep[e.g.,][]{dek92,yun95,han95,how01}. However, some detailed hydrodynamical calculations of the common envelope phase suggest that the value of $\alpha_{CE}$ is lower ($\sim$$\,$0.3--0.6; e.g., \citealp{ibe93,taa00}).  We should note that detailed hydrodynamical calculations cannot follow the CE phase completely through envelope ejection as yet, so any estimates of $\alpha_{CE}$ are based on extrapolations from these calculations, and should be regarded as suggestive rather than definitive.  The efficiency of the transfer of orbital energy into the common envelope remains a very uncertain quantity. 

In this paper, therefore, we quantitatively study the dependence of the formation rate of CVs with brown dwarf secondaries on the value of $\alpha_{CE}$.  In doing so, a question of particular interest is whether there is a minimum $\alpha_{CE}$ below which CVs with BD secondaries are {\it not} able to be formed.  We note that \citet{kol93} calculated model populations of secularly-evolved CVs using two different values of the efficiency parameter, $\alpha_{CE}$ = 1 and $\alpha_{CE}$ = 0.3, and found little difference in the model populations.  However, his study did not consider CVs that contained substellar secondaries at birth.  In addition, his study did not consider values of $\alpha_{CE}$ less than 0.3.  

To the best of our knowledge, the only hydrodynamical calculations of the spiral-in of a BD into a giant envelope are those of \citet{liv82}, \citet{liv83,liv84} and \citet{sok84}.  For the majority of their calculations, the authors investigate whether the secondary gains or loses mass during the spiral-in, and how this affects the outcome of the CE process.  Such an investigation is beyond the scope of the present paper.  In those calculations where the authors assumed the secondary mass did not change, they concluded that the spiral-in would lead to merger in most cases.  However, the calculations by Livio and Soker were one-dimensional, and thus crude compared with current three-dimensional hydrodynamical calculations \citep[cf.][]{ibe93,taa00}.

In the following section, we describe the method used to model the present-day formation of CVs with BD secondaries.  In section 3, we present results for the orbital period distribution of CVs forming with BD secondaries at the present time.  We also describe the results from our investigation of the dependence of the formation rate of CVs with BD secondaries on the assumed value of $\alpha_{CE}$.  In section 4, we present a discussion of the results and a comparison with observations.  We close with a summary of our main conclusions.

\section{Method}

The population synthesis code developed by \citet{pol88,pol96} to model the formation of CVs has been modified for the purposes of this study.  For a complete description of the code, we refer the reader to the original paper \citep{pol96}, and here only discuss the modifications made.  

(1) The code developed by \citet{pol88,pol96} employs a Jacobian technique to calculate the present-day CV birthrate and to map the ZAMS binary progenitor distributions to the ZACV distributions.  The Jacobian approach provides a completely analytic framework from which to calculate theoretical statistics of close binary systems from the parent population of ZAMS binaries.  However, in practice, it is difficult to test the effects of different values of certain input parameters (such as $\alpha_{CE}$ or the treatment of magnetic braking) on the computed ZACV population using the Jacobian code.  A Monte Carlo technique was used by \citet{dek92} to compute the CV birthrate.  He used identical or slightly modified physics as in \citet{pol88,pol96} and found substantially the same results.  A distinct advantage of the Monte Carlo approach over the Jacobian approach is that the Monte Carlo approach allows input parameters to be easily changed, and therefore lends itself nicely to calculating model populations for a wide range of input parameters.  Since \citet{dek92}, several other authors have used a Monte Carlo technique in population synthesis calculations of CVs \citep[e.g.,][]{dis94,han95,how01}.  We have therefore developed a Monte Carlo version of the Jacobian code, and have used the Monte Carlo code in the present study.  The Monte Carlo code contains the same physics as the Jacobian code, except as noted below, and it has been calibrated against the Jacobian code.

(2) The birthrate calculations of \citet{pol96} were modified to include a more realistic treatment of the radius of the BD as a function of mass and age. This treatment comes from cooling curves of BDs produced by I. Baraffe and co-workers \citep[Baraffe 2003, priv.\ comm.]{bar98}.  We show some representative curves in Figure 1.    The models in Figure 1 and similar models \citep[e.g.,][]{bur97} indicate that BDs, as they cool, approach a radius that is $\sim$ 0.1 $\Rsun$.  However, the cooling timescale to reach this radius can be as long as 10$^9$ years for the most massive BDs, and the radius can decrease by a factor of four.  Thus, unlike the calculations of \citet{pol96}, which assumed that the contraction phase to the main sequence for stars was short enough to be ignored, the cooling phase in BDs cannot be similarly ignored.  To avoid any ambiguity, we note that the term ``brown dwarf" is used in this paper to denote an object whose mass is between 0.013 and 0.075 $\Msun$ \citep[cf.][]{cb00}.  The lowest mass secondary considered in this paper is in accord with this definition.

(3) In many population synthesis studies of CVs, the parent ZAMS binaries are assumed to be distributed over primary mass (M$\rm_p$), mass ratio (q, defined as the ratio of the secondary mass to the primary mass), and orbital period (P) according to the following prescription:

\begin{equation}
   N(M_p,q,P)\;dM_p\,dq\,dP = f(M{_p})\,g(q)\,h(P)\;dM_p\,dq\,dP\,
\end{equation}
\linebreak
where f(M$\rm_p$), g(q) and h(P) are distributions over the respective orbital parameters \citep[e.g.,][]{pol88,pol96,dek92,dis94,han95}.   In the present Monte Carlo code, the distribution of primary masses, f(M$\rm_p$), is chosen from a \citet{mil79} initial mass function (IMF).  The distribution of progenitor orbital periods, h(P), is assumed to be flat in log P \citep[cf.][]{abt83}.  Both of these choices are the same as those in \citet{pol88,pol96}.  

For a given primary mass, the secondary mass is chosen from the distribution, i(M$\rm_s$) = M$\rm_p$$\,$g(q).  Therefore, the distribution of secondary masses is strongly influenced by the choice of g(q), the initial mass ratio distribution (IMRD) for the ZAMS progenitor binaries.  The IMRD chosen by \citet{pol88,pol96} was g(q) $\propto$ q.  This gives a distribution of initial mass ratios that is strongly peaked toward q = 1 (i.e., equal primary and secondary masses).  Observations of ZAMS spectroscopic binaries done after Politano's original work \citep{pol88} suggest that the IMRD is not very strongly peaked toward q = 1, but is flatter (i.e., g(q) $\propto$ q$\rm^a$, where a is very small; \citealp{duq91,gol03}).  For the calculation of the present-day orbital period distribution of ZACVs with BD secondaries, we have adopted an IMRD, g(q) dq = 1 dq.  We note though that both \citet{duq91} and \citet{gol03} point out that there are large uncertainties in their derivation of the IMRD for very small mass ratios (q $\la$ 0.20).  These uncertainties are due to the difficulty in observing the lines of the secondary star, since the light is dominated by the primary.  Thus, we emphasize that the IMRD for extreme mass ratio binaries remains very poorly constrained and our choice, g(q) = 1, is by no means a definitive one.  We further emphasize that, as in \citet{pol88,pol96}, we have assumed that the IMRD for binaries containing a ZAMS star and a BD {\it is the same} as the IMRD for binaries containing two ZAMS stars.  This is an important assumption whose validity is discussed later in section 4.2.3. 

In our Monte Carlo calculations, we begin with 10$^8$ ZAMS binaries.  This is sufficient to give a statistically significant sample of present-day ZACVs ($\sim$10$^4$ systems).   The age of the Galaxy is assumed to be 10$^{10}$ yrs. To reproduce the Miller-Scalo IMF, we used \citeauthor{egg01}'s \citeyearpar{egg01} Monte Carlo representation of it:

\begin{equation}
   M_p(u) = 0.19u[(1 - u)^{3/4} + 0.032(1 - u)^{1/4}]^{-1}\,
\end{equation}
\linebreak
where u is a random number uniformly distributed between 0 and 1.    As we are only concerned with the relative fraction of the total ZACV population that forms with BD secondaries, normalization of the above distributions has been ignored.

\section{Results}

The results of the population synthesis calculations are shown in Figures 2 and 3.  In Figure 2, the present-day orbital period distribution of ZACVs is shown for the assumed choices, $\alpha_{CE}$ = 1 and g(q) = 1.  The subset of ZACVs that contain BD secondaries is shown in the dashed histogram.  Our model predicts that CVs that are presently forming with BD secondaries occupy an orbital period range from 46 min to 2.5 hrs.  We further predict that 18\% of CVs at the present time are forming with BD secondaries.  Of these, the majority ($\sim$ 80\%) form below the orbital period minimum (78 min), with the remainder forming a tail reaching within the stellar domain.  Thus, our model predicts that a {\it significant fraction}, $\sim$15\%, of CVs are forming at orbital periods {\it shorter} than 78 minutes at the present time.

In Figure 3, the fraction of present-day ZACVs that contain BD secondaries is plotted as a function of  $\alpha_{CE}$, the common envelope efficiency parameter, for 3 different IMRDs.  The IMRDs chosen are: g(q) $\propto$ q, g(q) $\propto$ 1, and g(q) $\propto$ q$^{-0.9}$. Since the total number of ZACVs is sensitive to the choice of IMRD \citep[cf.][]{dek92,pol94}, it may be true that the fraction of CVs that form with BD secondaries also is sensitive to this choice.  In light of the aforementioned lack of strong observational constraints on the IMRD, we chose functional forms for g(q) that range from a distribution that strongly favors the formation of extreme mass ratio ZAMS binaries (g(q) $\propto$ q$^{-0.9}$), to one that strongly favors the formation of ZAMS binaries that contain equal mass components (g(q) $\propto$ q). 

Not surprisingly, we find that the fraction of ZACVs with BD secondaries increases as the IMRD favors more and more extreme mass ratio progenitor binaries.  For $\alpha_{CE}$ = 1, the value used in the calculation of the orbital period distribution in Figure 2, this fraction ranges from 8\%, for g(q) $\propto$ q, to slightly greater than one-third (34\%), for g(q) $\propto$ q$^{-0.9}$.  For a given IMRD, the fraction of ZACVs with BD secondaries is relatively constant from $\alpha_{CE}$ = 1.0 to $\alpha_{CE}$ = 0.3, and then begins to decrease for $\alpha_{CE} <$ 0.3.  However, we find that {\it even for very small values} of the common envelope efficiency parameter ($\alpha_{CE}$ = 0.1), the fraction of CVs that form with BD secondaries is {\it not} zero.  Rather, for $\alpha_{CE}$ = 0.1, this fraction ranges from $\sim$ 3\% to $\sim$ 18\%, depending on the IMRD chosen.

\section{Discussion}

\subsection {Comparison with Observations}

Our models predict that the majority of CVs that form with BD secondaries
populate an orbital period interval, 46 min $<$ P$_{\rm orb}$ $<$ 78 min, that is {\it not} populated by CVs that formed with ZAMS secondary stars.  This result, calculated using up-to-date BD models, confirms the previous result by \citet{pol96} that ultra-short period CVs (P$_{\rm orb} \la 80$ min) can be formed if the secondary is substellar at birth.  Such systems therefore may provide a viable model for the observed ultrashort-period CVs, V485 Cen and RX 2329+06.  We discuss this possibility below.  In addition, there are a handful of CVs {\it above} the period minimum, in the orbital period range $\sim$ 78-90 mins, that are strongly suspected of containing BD secondaries.  Since, ZACVs with BD secondaries extend in orbital period to 2.5 hours, we also discuss below whether these handful of suspected BDCV systems may be CVs that formed with BD secondaries. 

We recognize that we are comparing a present-day orbital period distribution of ZACVs with BD secondaries with the observed, present-day orbital period distribution of CVs.  Clearly, a more correct comparison would be between a secularly-evolved BDCV orbital period distribution and the observed one.  However, the mass transfer rates in this particular sub-population of CVs are so low ($\sim$10$^{-11}$ $\Msun$ yr$^{-1}$; \citealp{kol99}) that we do not expect the secularly-evolved distribution to be vastly different than the zero-age one.  For example, for the secular evolution of a 0.07 $\Msun$ BD and a 0.6 $\Msun$ WD, \citet{kol99} found that it took 1.5 Gyrs for the system to evolve from 54 min to 78 min.  Therefore, we do not anticipate that subsequent secular evolution of ZACVs with BD secondaries will evacuate the orbital period regime $<$ 78 min.  Nevertheless, our estimates of the fraction of ZACVs with BD secondaries should be regarded as upper limits.  Plans to extend the present calculation to include subsequent secular evolution are in progress.

\subsubsection {V485 Cen and RX2329+06}
As mentioned earlier, V485 Cen and RX2329+06 are unusual CVs because their orbital periods of 59 min and 64 min, respectively, are significantly shorter than the orbital period minimum for CVs.  Suggestions as to why V485 Cen and RX2329+06 have periods shorter than the period minimum have focused on the possibility that the secondary star in these systems had undergone significant nuclear evolution in its core prior to mass transfer \citep{aug96, sch02, tho02, uem02}.  Theoretical calculations by \citet{sch02} indicate that a CV containing a secondary core in which 95\% of the hydrogen has been burned into helium prior to mass transfer could reach a minimum period $\la$ 1 hr.  A similar calculation by \citet{tho02} found comparable results \citep[see also][]{pyl88,pod02}.

While ZACVs with BD secondaries can account for the ultra-short orbital periods in V485 Cen and RX2329+06, recent spectroscopic observations of RX2329+06 by \citet{tho02} have determined an unusually high temperature of $\sim$$\,$4000 K for the secondary.  Howell (2002, priv. comm.) has also found a similar temperature for the secondary in V485 Cen.  If these temperature determinations are correct, then this presents serious problems for our model.  Even the youngest, hottest BDs do not have surface temperatures as high as 4000 K (I. Baraffe 2002, priv. comm.; see also \citealp{cb00}).  It is possible that the close proximity of a hot white dwarf could cause significant heating of the BD atmosphere.  Hauschildt (2002, priv. comm.) has performed preliminary calculations of the irradiation of a gas giant planet by a 15,000 K white dwarf at a distance of 1 AU.  He did not find a significant increase in the temperature of the planet.  However, the orbital separations in V485 Cen and RX2329+06 are much smaller, of order a few tenths of a solar radius.  \citet{har03} have recently modeled the light curve of EF Eri, a CV with an orbital period of 81 min that is strongly suspected of containing a substellar secondary (see section 4.1.2. below).  They infer from their models that the secondary star in EF Eri is very cool (T$_{\rm eff} \sim$ 900 K) and is irradiated by the WD.  They estimate the irradiated side of the secondary has a temperature, T$_{\rm eff} \la$ 1600 K.  However, they note that proper modeling of the light curve of EF Eri will require further work, such as the incorporation of BD atmospheres into the light curve program used by them.  At the moment, therefore, it is not known whether the conditions that exist in V485 Cen and RX 2329+06 could provide enough heating of the BD atmosphere to account for the observed temperatures. 

Other observed features in V485 Cen and RX2329+06 also favor the evolved secondary model.  Both systems have HeI $\lambda$6678/H$\alpha$ ratios that are higher than what is seen in typical SU UMa systems \citep{aug96,tho02}.  This suggests that the accreted material is enriched in helium.  In addition, the observational estimates of the secondary mass in V485 Cen (M$_{\rm sec} \ga $ 0.14 $\Msun$; \citealp{aug96}) and RX2329+06 (M$_{\rm sec} \sim$ 0.13 $\Msun$, assuming a typical mass of 0.6 $\Msun$ for the WD; \citealp{tho02}) are not consistent with a substellar secondary.

\subsubsection{CVs above the Period Minimum Suspected to Contain BD Secondaries}

In a recent paper, \citet{lit03} reviewed the observational evidence for BD secondaries in CVs.  They conclude that there is signficant indirect evidence for BD secondary stars in 6 CVs, all of which have orbital periods greater than 78 minutes.  These systems are DI UMa (78.6 min), EF Eri (81.0 min), WZ Sge (81.6 min), AL Com (81.6 min), EG Cnc (86.3 min), and VY Aqr (90.8 min).  As noted in the Introduction, according to the standard model of CV secular evolution \citep[e.g.,][]{kin88}, as the secondary star transfers mass to the WD, the orbital period of the system shrinks until, at a certain point, the secondary's thermal timescale becomes longer than the mass transfer timescale \citep{pac81,rap82}.  This point roughly coincides with the secondary having lost so much mass that it has become substellar \citep[cf.][]{how97}.  Further mass transfer then results in an increase in the orbital period of the system \citep{pac81,rap82}.  From Figure 1, we see that the orbital period distribution of present-day ZACVs containing BD secondaries extends above 78 minutes to 2.5 hours. Therefore, at the present time, we cannot unambiguously say whether the above 6 systems are CVs that contain secondaries that were {\it born} substellar or {\it became} substellar as a result of mass loss during secular evolution.  Below we compare the likelihood of each.

Theoretical studies of the intrinsic (i.e., actual, not observed), current CV population predict that, according to standard CV secular evolution, 99\% of the current population of CVs have orbital periods less than 3 hrs, and of these, 60-70\% contain secondaries that have lost enough mass to become substellar \citep{kol93,how97,how01}.  Under the assumption of a constant CV birthrate throughout the Galaxy's history \citep[e.g.,][]{kol93}, our calculations predict an upper limit to the percentage of CVs in the period range 78 min--3 hr, whose secondaries were born substellar, of 18\%.  This assumes all CVs that form below 78 minutes quickly evolve to orbital periods greater than 78 minutes (as discussed above, this is clearly an unrealistic assumption but we use it as an extreme case to calculate an upper limit).  Therefore, assuming standard secular evolution, it is at least $\sim$ 4 times more likely that CVs in the period range 78 min--3 hr contain secondaries that became substellar due to mass loss rather than were born substellar.  

We note that recent observations of the rotation rates of low-mass stars in open clusters \citep{tern00} have raised some concerns about the validity of the standard model of CV secular evolution.  An empirical angular momentum loss prescription inferred from these observations \citep{and03} suggests that the strength of the mean magnetic braking in CVs may be on the order of 100 times less than that assumed in the standard model for CVs above the period gap (P$_{\rm orb}$ $>$ 3 hrs; cf. \citealp{kin88}). Such a drastic reduction in the mean magnetic braking would imply proportionately longer lifetimes for CVs above the gap.  If the magnetic braking above the gap is indeed as low as suggested by \citet{and03}, then \citet{kin02} have postulated that the 78 min cutoff in the orbital period distribution of CVs may be a feature dictated by the age of the Galaxy.  That is, 78 minutes may simply be the shortest orbital period achievable by a CV within the age of the Galaxy, given the much longer evolutionary timescales above the gap due to reduced magnetic braking.  If true, this would drastically reduce the large population of CVs (predicted by the standard model) whose secondaries became substellar as a result of mass loss, and thereby increase the likelihood that systems like WZ Sge contained substellar secondaries at birth.

The rotation rate data from open clusters also indicate no significant reduction in magnetic braking once a star becomes fully convective \citep{tern00}, as assumed in the standard model \citep[e.g.,][]{rap83}.  Thus, magnetic braking should also operate below the period gap, whereas in the standard model, the only angular momentum mechanism driving mass transfer below the gap is gravitational radiation \citep{pac81,rap82}.  \citet{and03} infer an angular momentum loss rate below the gap that is $\sim$ 1.5 times higher than gravitational radiation alone.  Such higher mass transfer rates below the gap could completely exhaust the mass of the secondary within the age of the Galaxy, and thus significantly reduce the number of secondaries that became substellar as a result of mass loss \citep[cf.][]{pat98}.  It should be noted that the empirical angular momentum loss prescription by \citet{and03} does not provide a satisfactory explanation for the existence of the period gap in CVs, or the observed brightnesses of CVs above the gap.

\subsection{Where are All the CVs with Orbital Periods less than 78 mins?}

There are approximately 500 CVs with known or suspected orbital periods in the \citet{rit03} catalog.  Regardless of whether V485 Cen or RX2329+06 contained an evolved secondary or a substellar secondary at birth, the observed number of CVs in the period range 46 min-78 min is still considerably less than predicted by our models of ZACVs with BD secondaries (15\% or $\sim$ 75 CVs) for an IMRD, g(q) = 1, and $\alpha_{CE}$ = 1.  Even with the recognition that 15\% is an upper limit, the discrepancy between our predicted number of CVs with orbital periods less than 78 mins and the observed number is significant.  In the sections below, we discuss possible reasons for the cause of this discrepancy.

\subsubsection{Observational Selection Effects}

The mass transfer rates in ZACVs with BD secondaries are predicted to be low ($\sim$ 10$^{-11}$ $\Msun$ yr$^{-1}$; \citealp{kol99}), therefore these systems should be faint (m$_v \ga$ 20; \citealp{pol98}) and difficult to observe.  Consequently, it is possible that these systems exist, but are too faint to be observed currently in quiescence.  Indeed, the 6 systems suspected of containing BD secondaries discussed above were discovered as a result of their outbursts.  The recurrence times betweeen outbursts in these six systems is longer than most dwarf novae \citep[cf.][]{war95}, and this may be a property of CVs that contain substellar secondaries.  Presumably, the longer recurrence times are a result of the comparatively lower mass transfer rates in these systems than typical dwarf novae.  \citet{kol99} calculated evolutionary tracks for CVs with initial secondary masses between 0.04 $\Msun$ and 0.21 $\Msun$.  They computed mass transfer rate as a function of orbital period, and found that the evolutionary tracks merge into a common track for orbital periods above approximately 70 minutes, {\it regardless} of whether the secondary star was initially substellar or became substellar as a result of mass loss (cf. Fig. 1 in \citealp{kol99}; we note that the BD models used in \citet{kol99} are very similar to the ones used in the current paper.)  Their tracks also show that a CV that formed with a substellar secondary below 70 mins (0.05 $\Msun \la$ M$_{\rm sec} <$ 0.075 $\Msun$) has a mass transfer rate higher than its evolved counterpart above the period minimum whose secondary became substellar as a result of mass loss (again, cf. Figure 1 in \citealp{kol99}).  Thus, although observational selection effects cannot be ruled out, if we are observing outbursts in systems with presumably substellar secondaries above the period minimum (e.g., WZ Sge, AL Com, etc.), then it is difficult to understand why we would not be able to observe outbursts in systems formed below the period minimum, if the mass transfer rates are the same or higher.

\subsubsection {Common Envelope Efficiency Parameter for Very Low Mass Secondaries}

Our results indicate that CVs will form with BD secondaries unless $\alpha_{CE}$ is {\it extremely} small ($<$ 0.1).  At first glance, this is a surprising result.  However, in the population synthesis code used in these calculations, the orbital separation at the end of the CE phase is only limited by the condition that the secondary does not fill its Roche lobe during the spiral-in process.  This effectively allows spiral-in for such small objects to an orbital separation of $\sim$ a few tenths of a solar radius.  Thus, the decrease in the orbital energy reservoir due to such small secondary masses is offset by the very small orbital separations these systems may achieve upon exit of the CE phase.  Therefore, only if the transfer of orbital energy into the envelope is extremely inefficient will merger occur.  

Such extremely small values of $\alpha_{CE}$ for the entire population of CV progenitors would produce a total number of CVs in the Galaxy that would be difficult to reconcile with even the lowest estimates of the CV space density 
($\sim$$\,$10$^{-6}\,$pc$^{-3}$; \citealp{pat84}).  In addition, estimates of the efficiency parameter extrapolated from detailed hydrodynamical calculations of the CE phase have typically indicated $\alpha_{CE}$ $\ga$ 0.3 \citep[e.g.,][]{ibe93,taa00}. 

It may be that $\alpha_{CE}$ is a function of the secondary mass.  In particular, if $\alpha_{CE}$ decreases sharply as the secondary mass approaches the substellar regime, this could account for the lack of observed CVs below 78 minutes.  For example, it may be true that the luminosity generated by the drag torque on the secondary becomes comparable to the luminosity of the giant star for very low mass secondaries.   In this case, merger cannot be avoided since the time scale for the generation of frictional energy would be too long to dynamically expel the envelope (R. Taam 2002, priv. comm.; see also \citealp{ibe93} and references therein).  

A second possibility is that $\alpha_{CE}$ is different for CE evolution involving primaries that fill their Roche lobes on their initial ascent of the giant branch (i.e., before He ignition) than for primaries that fill their Roche lobes as AGB stars.  There have been several detailed hydrodynamical calculations investigating the spiral-in of a low mass secondary star into an AGB primary \citep[cf.][]{ibe93}.  However, there has been only one detailed study of the spiral-in of a low-mass secondary into an {\it RGB} primary \citep{san00}.  It appears from this study that $\alpha_{CE}$ could be different for CE evolution involving an RGB primary than for an AGB primary.  However, further detailed study of this scenario is necessary before any firm claim can be made.  Population synthesis calculations of the formation of CVs that include a variable $\alpha_{CE}$ are currently underway \citep{pol03}.   

Finally, although our results indicate that there is a sufficient energy reservoir in the orbit to unbind the envelope even if $\alpha_{CE}$ is quite small, Taam and co-workers have noted that the drag torque on the secondary must also approach zero, or else the secondary will continue to spiral in towards the core.  They find that this dynamical constraint on the drag torque requires the presence of steep density gradients within the common envelope \citep{yor95,ter96}.  To the best of our knowledge, no population synthesis code for CVs (including our own) includes this dynamical constraint on the drag torque in its treatment of the common envelope phase.

\subsubsection {IMF and Mass Ratio Distribution for Binaries with Substellar Companions}

Several studies have shown that the daughter ZACV population is strongly dependent on the assumed IMRD chosen for the parent ZAMS binary population \citep[e.g.,][]{dek92,pol94}.  As evident in Fig. 3, the population of ZACVs with BD secondaries is similarly strongly affected by the assumed choice of the progenitor IMRD.  In the present study, we have assumed that the IMRD for binaries containing a ZAMS star and a BD is the {\it same} as the IMRD for binaries containing two ZAMS stars.  As described below, preliminary results for the BD mass function and the frequency of BDs in binaries suggest that this assumption likely is not valid. 

Observational searches for BDs in young clusters \citep{mart98,mart00,bou98,ham99}, in star forming clusters \citep[e.g.,][]{luh00,luhet00,bou01}, and in the field \citep{rei99} suggest that the mass function for single BDs is approximately flat or rises slowly from the substellar regime to $\sim$ 0.6 $\Msun$, and then rolls over into a power law from $\sim$ 1 $\Msun$ to higher masses.  Estimates of the number density of nearby BDs from surveys such as the Deep Near-Infrared Survey (DENIS; \citealp{del99}) and the Two Micron All-Sky Survey (2MASS; \citealp{kir99,bur99}) are consistent with such a BD mass function, although clearly we are dealing with small number statistics. 

Investigation of the mass function of BDs in binaries is an even more recent undertaking.  Observational data in the past few years have allowed some preliminary statements to be made.  These data indicate that BD companions to F-M stars in ``close'' orbits ($\la$ 3 AU) are rare (the so-called ``brown dwarf desert;'' cf. \citealp{marc00}).   Recent observations by \citet{giz01} indicate that no such desert exists for BD companions to F-M stars in wide orbits ($\ga$ 1000 AU).  

From the present calculations, we find that CVs that are currently forming with BD secondaries have evolved from a parent, ZAMS population of binaries with orbital separations $<$ 3 AU and primary star masses in the range $\sim$ 1-10 $\Msun$.  In Figure 4, we present histograms of the number of progenitors as a function of (a) the progenitor ZAMS primary mass and (b) the progenitor ZAMS orbital separation, respectively.  Inspection of Figure 4a indicates that $\sim$ 75\% of the primaries in the parent, ZAMS population have masses less than $\sim$ 1.6 $\Msun$, and thus have spectral types F or later.  Inspection of Figure 4b shows that the vast majority of the progenitors have orbital separations less than 1 AU.  Further, (although this cannot be discerned in the figure), the largest orbital separation in the progenitor population is 2.98 AU.  Interestingly, these ranges in orbital separation and primary mass for approximately three-fourths of the progenitors of ZACVs with BD secondaries lie directly within the brown dwarf desert.  It is possible, therefore, that the lack of observed CVs with orbital periods less than 78 minutes may be due to a lack of progenitor binaries from which to draw.   Although the existence of the brown dwarf desert seems fairly well established, we believe that further detailed observations of substellar objects in close binaries are needed before any firm conclusions can be drawn.  Furthermore, our understanding of $\alpha_{CE}$ for CE evolution involving very low-mass secondaries needs to improve before we can separate its effect on the formation of ZACVs with BD secondaries from the effect of the IMRD.

The result that all progenitors of ZACVs with BD secondaries have ZAMS orbital separations less than 3 AU is nevertheless quite significant, and underlines the potential importance of observations of CVs with ultra-short orbital periods.   Assuming the effect of $\alpha_{CE}$ on the results can be isolated, then it may be possible to use observations of CVs with orbital periods less than 78 mins, along with the models presented in this paper, to provide useful information about the progenitor population of ZAMS + BD binaries.  For example, by comparing our models with these observations, we may be able to place constraints on the BD mass function and/or IMRD.  These constraints would nicely complement, and possibly augment, our understanding of the BD mass function/IMRD in MS binaries from observations of these systems in clusters and in the field, for example.  In addition, a firm null result for the detection of CVs with BD secondaries and orbital periods less than 78 min has the potential to provide strong, independent evidence for the existence of the brown dwarf desert.    

We close this section with a few speculative remarks regarding the formation of BDs in binaries.  According to standard stellar evolution theory, primary stars with masses $\la$ 1$\,$$\Msun$ will not evolve off of the main sequence within the age of the Galaxy.  One might therefore expect to see a relative increase in the number of BD-MS star binaries with primary masses below 1 $\Msun$.  However, the brown dwarf desert extends in primary mass well below 1 $\Msun$ to M-type primaries.  Secondly, an observational point that has not received as much attention as the brown dwarf desert is that there is {\it no} observed scarcity of L dwarf--L dwarf binaries (cf. \citealp{rei01,cha02} and references therein).   These statements suggest that there is something different about the frequency or formation rate of BD--MS star binaries compared with MS-MS binaries or BD--BD binaries.  Much attention has been paid to the BD mass function in binaries.  We suggest that perhaps the more important distribution is the mass ratio distribution in binaries containing BDs.  The existing observational data are also consistent with the following statement:  BDs in binaries with approximately equal mass ratios are not rare, while BDs in binaries with extreme mass ratios are, at least at separations $\la$ 3 AU.  This raises some interesting questions:  Is the frequency of BDs in binaries strongly affected by the mass ratio?  Is the mass ratio distribution in extreme mass ratio binaries independent of the total mass of the binary?  For example, is there a corresponding relative scarcity of ZAMS binaries with extreme mass ratios compared to those with nearly equal mass ratios?  \citet{gol03} have recently derived the mass ratio distribution for a sample of 129 spectroscopic binaries with orbital periods less than 2500 days.  They find that there is a sharp drop in the mass ratio distribution for mass ratios less than $\sim$ 0.2 (again, secondary mass/primary mass).  However, they caution that there are large uncertainties in their derivation for such small ratios, and they are hesitant to ascribe a statistical significance to this finding.  We believe each of these questions merit further study.

\section{Summary and Conclusions}

We summarize our key findings and conclusions below:
 
(1) CVs that are presently forming with BD secondaries are predicted to occupy an orbital period range of 46 min to 2.5 hrs.  This result is in substantial agreement with the previous result by \citet{pol96}.  For the assumed choices of $\alpha_{CE}$ = 1 and a mass ratio distribution in the ZAMS progenitor binaries, g(q) dq = 1 dq, we predict that 18\% of all present-day ZACVs contain secondaries that are BDs.  Of this 18\%, the majority ($\sim$ 80\%) form below the orbital period minimum (78 min), with the remainder forming a tail reaching within the stellar domain.  Thus, our model predicts that a {\it significant fraction}, $\sim$15\%, of CVs are forming with orbital periods {\it shorter} than 78 minutes at the present time.  

(2) CVs that form with BD secondaries could provide a viable model to explain the ultra-short periods in the unusual CVs, V485 Cen (P$_{\rm orb}$ = 59 min) and RX2329+06 (P$_{\rm orb}$ = 64 min).  However, because of the unusually high temperatures derived for the secondaries ($\sim$ 4000 K) in these systems (\citealp{tho02}; Howell 2002, priv. comm.), it appears unlikely that these systems contain a BD secondary, unless irradiation of the BD by the nearby WD has a significant effect on the surface temperature of the BD.  These high temperatures for the secondary are more easily explained with a model in which the secondaries in V485 Cen and RX2329+06 are the cores of evolved MS stars \citep{sch02,tho02,uem02}.  This model is also in better agreement with the estimated secondary masses and enhanced He/H ratio observed in these systems.

(3) CVs that form with BD secondaries could also provide a model for six CVs above the period minimum, DI UMa, EF Eri, WZ Sge, AL Com, EG Cnc, and VY Aqr, for which there is signficant indirect evidence that the secondary stars in these systems are BDs \citep{lit03}.  These six systems have orbital periods between $\sim$ 78-90 min.  However, according to the standard model of CV secular evolution, it is at least four times more likely that the secondaries in these six systems became substellar as a result of mass loss, rather than were born substellar.  It is important to note though that the validity of the magnetic braking prescription used in the standard model has been recently called into question by observations of the rotation rates of low-mass stars in open clusters \citep{tern00, pin02}.  A prescription for magnetic braking inferred from the open cluster data \citep{and03} could imply a significant reduction in the number of CVs whose secondaries became substellar as a result of mass loss \citep[cf.][]{pat98,kin02}.  If true, this would correspondingly increase the likelihood that WZ Sge, AL Com, etc. are CVs that formed with BD secondaries.

(4) Assuming a constant CV birthrate throughout the Galaxy's history, the fraction of CVs that are presently forming below the period minimum according to our model, 15\%, represents an upper limit to the predicted fraction of {\it all} CVs with orbital periods less than 78 minutes at the present time.  Of the $\sim$ 500 CVs with known orbital periods, 15\% represents 75 systems.  The number of CVs with orbital periods less than 78 minutes is currently two, V485 Cen and RX 2329+06, and it is dubious whether these systems contained substellar secondaries at birth.  Thus, even with the recognition that 75 systems is an upper limit, the observed number of CVs in the period range 46 min-78 min is still {\it considerably less} than the number predicted by our models.  It does not seem likely that the discrepancy is due to observational selection effects.  The mass transfer rates and, hence, recurrence times in CVs with BD secondaries below the period minimum should be very similar to systems like WZ Sge and AL Com \citep[cf.][]{kol99}.  Consequently, it is difficult to understand why we would not have observed outbursts in them.  Ongoing searches for faint CVs in observational surveys such as 2MASS \citep{hoa02,wac03}, the Sloan Digital Sky Survey (SDSS; \citealp{szk02}) and the Faint Sky Variability Survey (FSVS; \citealp{how02}) may be able to eventually detect such systems in quiescence, assuming they exist.  

(5) We have examined the dependence of the present-day formation of CVs with BD secondaries on the assumed value of the common envelope efficiency parameter, $\alpha_{CE}$.  We find that even for values of $\alpha_{CE}$ as low as 0.1, CVs with BD secondaries are still able to be formed.  For $\alpha_{CE}$ = 0.1, our calculations predict that the present-day fraction of ZACVs with BD secondaries is 8\% for an flat IMRD, g(q) dq = 1 dq.  For an IMRD that strongly favors the formation of binaries with mass ratios near unity, g(q) $\propto$ q, this fraction is decreased to 3\%.  For an IMRD that strongly favors the formation of binaries with extreme mass ratios near, g(q) $\propto$ q$^{-0.9}$, this fraction is increased to 18\%.  A firm null result for the detection of CVs with BD secondaries having orbital periods less than 78 mins could therefore imply that the transfer of orbital energy into the common envelope is extremely inefficient for systems with very low-mass secondaries.  Further, as it is unlikely that the CE process is that inefficient for the entire CV population, such a null detection would also then imply that the efficiency of the CE process is a function of secondary mass.  If true, it would then seem extremely coincidental if $\alpha_{CE}$ decreases abruptly exactly at the substellar transition mass.  Rather, one would expect that the decrease is smoother and begins in the stellar domain.  Consequently, determining if there is a cutoff mass for the secondary, below which merger is inescapable, may have important implications for our theoretical understanding of the orbital period distribution in CVs below, and possibly within, the period gap.

(6)  Lastly, CVs that form with BD secondaries have evolved from progenitor ZAMS + BD binaries with primary masses in the range $\sim$ 1-10 $\Msun$ and with orbital separations $<$ 3 AU, for an assumed IMRD that is flat (g(q) = 1).  Furthermore, 75\% of these progenitor systems have primary masses $\la$ 1.6 $\Msun$, and therefore spectral types F or later.  The orbital parameter ranges in primary mass and orbital separation for these 75\% coincide with the so-called ``brown dwarf desert.''  Comparison of the models presented in this paper with observations of ultra-short period CVs may therefore be able to provide important information regarding the formation and distribution of BD + MS binaries with orbital separations less than 3 AU.

In summary, we believe that the study of the formation of CVs with BD secondaries can provide useful insights concerning our understanding of common envelope evolution in close binary systems with very low mass secondaries, and the distribution of BDs in MS binaries having orbital separations $<$ 3 AU.  We urge theorists who are performing detailed hydrodynamical calculations of the CE phase to investigate in more detail the spiral-in of very low-mass objects in order to determine if there is indeed a cutoff mass below which merger is unavoidable.  We are extremely grateful to the increasing numbers of observers involved in searches for faint CVs.  These searches ultimately hold the key to our understanding of the global CV population.  We urge observers involved in such searches, especially in the infrared, to be particularly vigilant regarding ultrashort-period CVs ($\sim$ 40-80 mins).  For the reasons discussed in this paper, these systems may provide important guidance for constructing models of CV formation and secular evolution.  This is particularly important at this time, as we reexamine, and in some cases abandon, long-held beliefs about the formation and secular evolution of CVs.

\acknowledgements {Warm thanks to Drs. T. Augusteijn, I. Baraffe, L. Bildsten, P. Hauschildt, D. Hoard, S. Howell, J.D. Kirkpatrick, U. Kolb, P. Lawrence, E. Sandquist, K. Schenker, P. Szkody, and R. Taam for very useful discussions about this work.  We especially thank Dr. I. Baraffe for providing the BD models.  Lastly, we thank an anonymous referee whose comments helped greatly to improve the paper and make it more accessible to a wider readership.  This work was funded in part by NASA grant NAG 5-8481 and NSF grant AST-0098742 to Arizona State University, and NSF grant AST-0328484 to Marquette University.}

\clearpage


\begin{figure}
\plotone{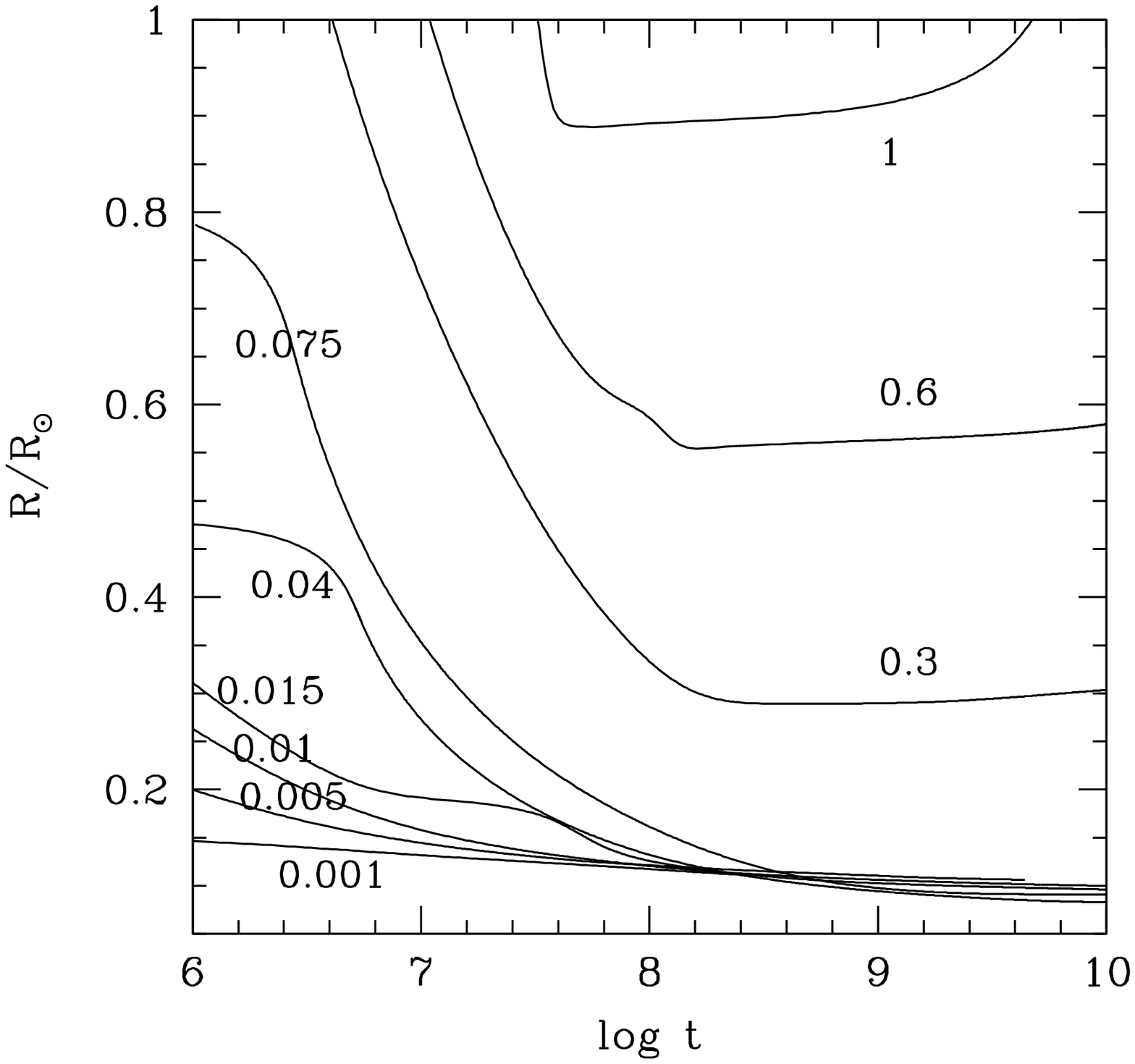}
\figcaption{Radius as a function of time for BDs and low mass stars of solar composition.  Numbers labeling curves indicate mass in solar masses.  The time is measured in years from the formation of the BD.  These cooling curves were taken from detailed model calculations by Baraffe and co-workers (Baraffe et al. 1998; Baraffe 2003, priv. comm.)}
\end{figure}

\clearpage

\begin{figure}
\plotone{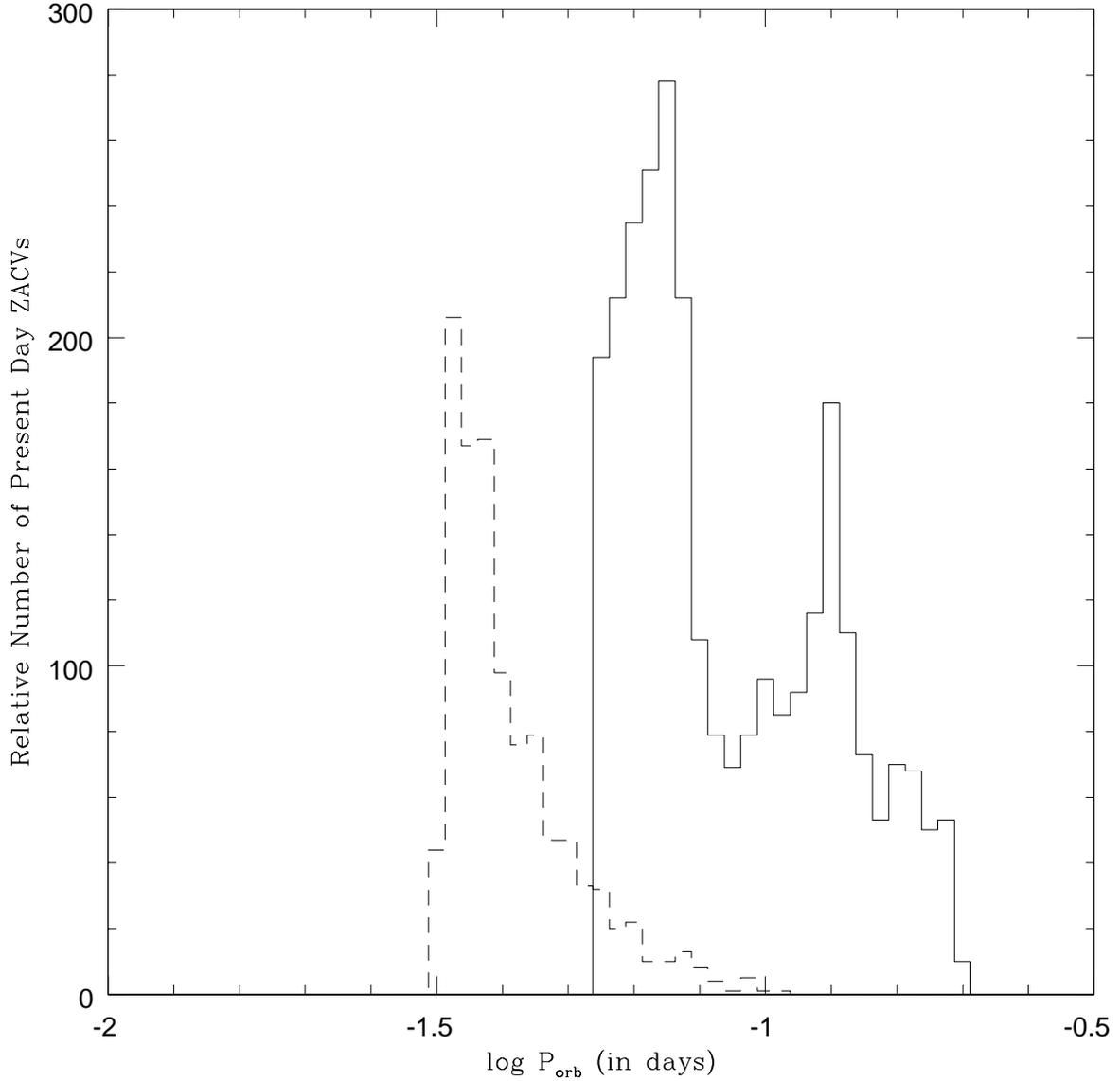}
\figcaption{The predicted distribution of orbital periods in present-day ZACVs for the assumed choices: $\alpha_{CE}$ = 1 and g(q) dq = 1 dq.  The distribution has been divided into two histograms.  The dashed histogram shows systems that contain BD secondaries.  The solid histogram shows systems that contain stellar secondaries.  The total distribution is the sum of these two histograms.}
\end{figure}

\begin{figure}
\plotone{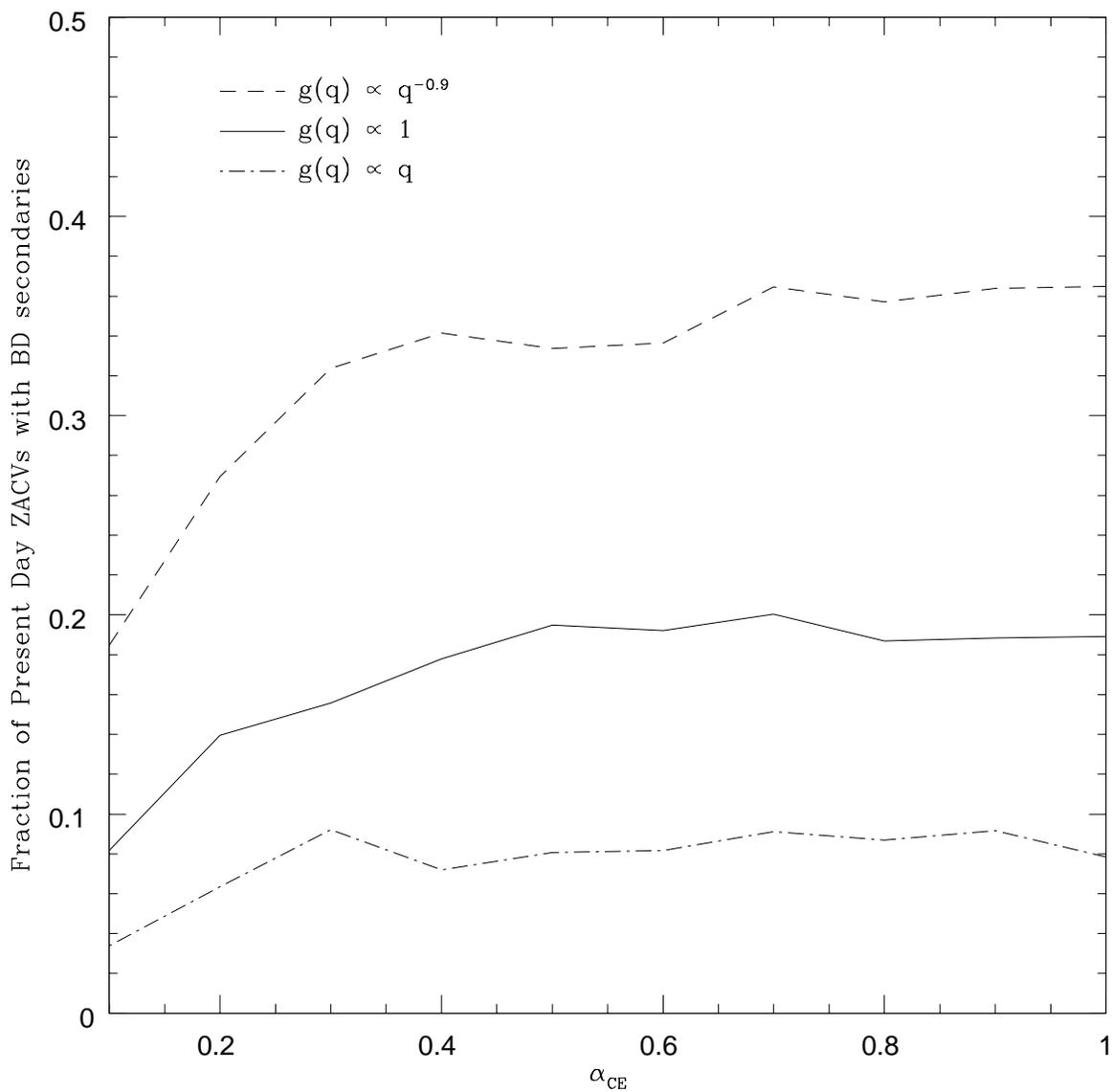}
\figcaption{The fraction of present-day ZACVs that contain a BD secondary as a function of $\alpha_{CE}$ for three different IMRDs.  The dashed line is for g(q) $\propto$ q$^{-0.9}$, the solid line for g(q) $\propto$ 1, and the dashed-dotted line for g(q) $\propto$ q.}
\end{figure}

\begin{figure}
\plottwo{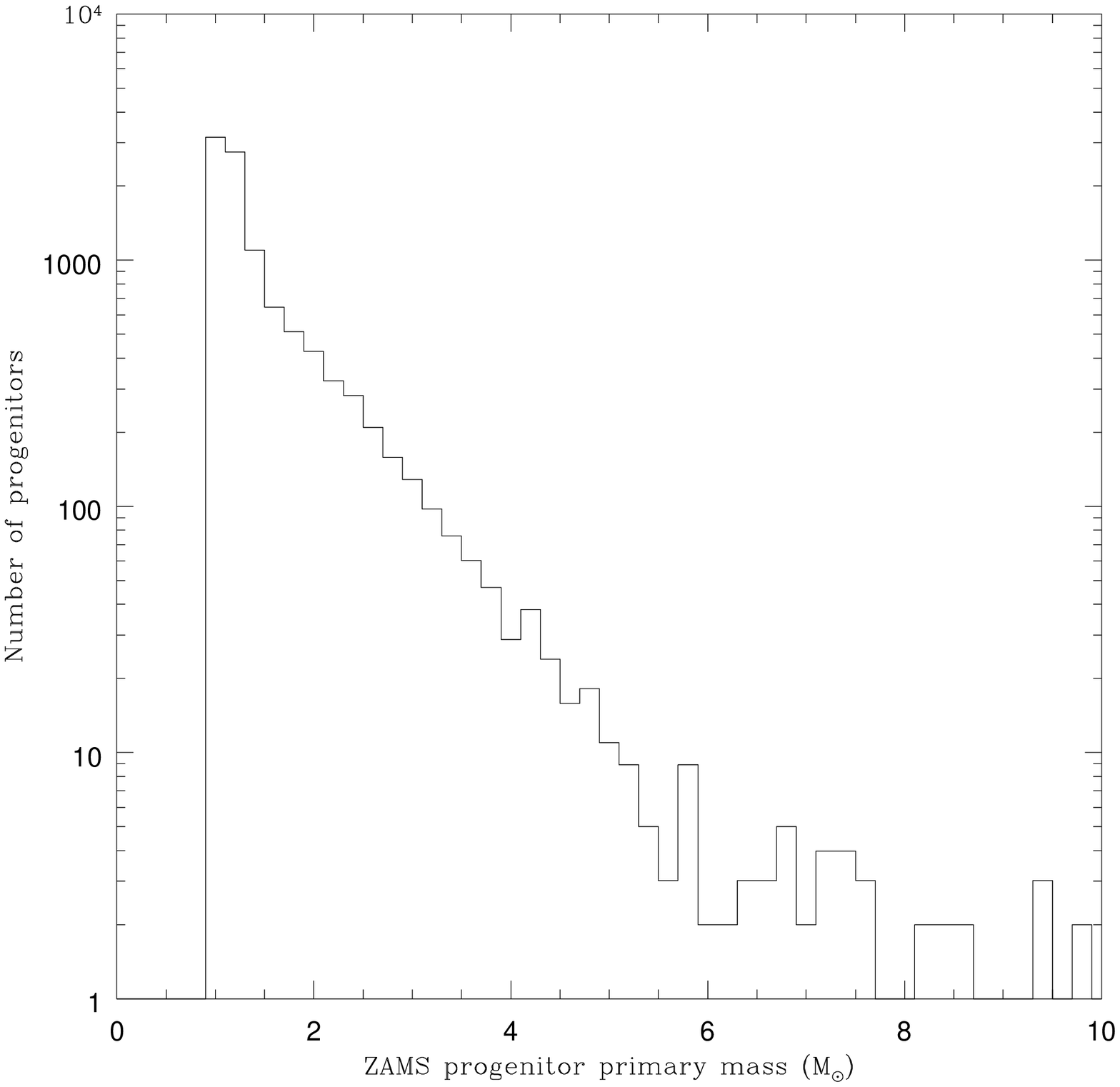}{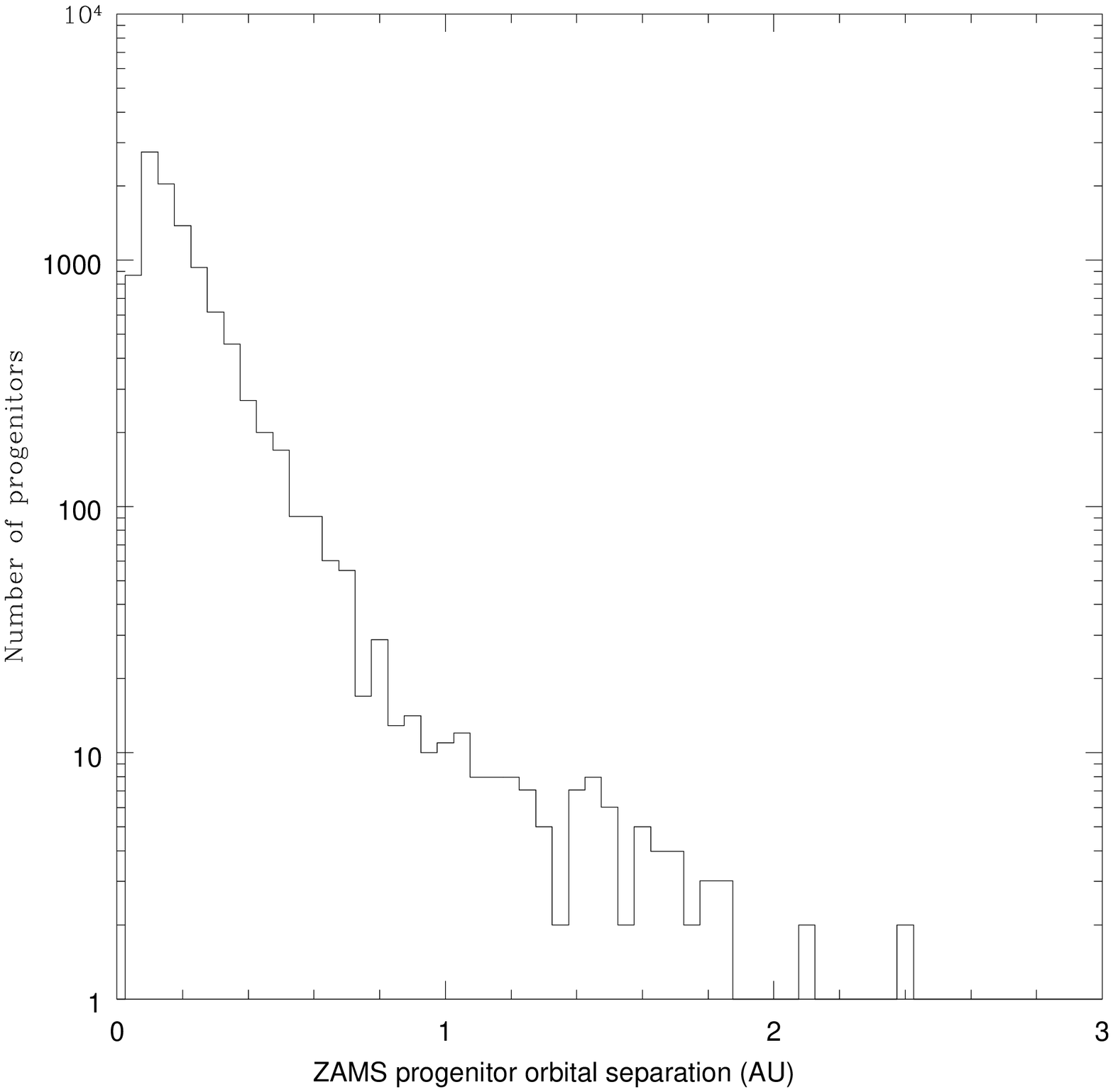}
\figcaption{(a) The distribution of ZAMS primary masses in the progenitors of CVs that are forming with BD secondaries at the present epoch.  (b) The distribution of ZAMS orbital separations in the progenitors of CVs that are forming with BD secondaries at the present epoch.  The binaries in these distributions are the parent population of the ZACV binaries shown in the dashed histogram (BD secondaries) in Figure 2.  (We note though that we increased the number of initial binaries in our Monte Carlo calculations from 10$^8$ to 10$^9$ to ensure reasonable statistics for the entire range of primary masses in the above distribution.)}
\end{figure}


\begin{thebibliography}{}

\bibitem [Abt(1983)] {abt83}
Abt, H.A. 1983, \araa, 21, 343
\bibitem [Allard et al.(1997)] {all97}
Allard, F., Hauschildt, P., Alexander, D.R, \& Starrfield, S., 1997, \araa, 35, 137
\bibitem [Andronov et al.(2003)] {and03}
Andronov, N., Pinsonneault, M., \& Sills, A., 2003, \apj, 582, 358
\bibitem [Augusteijn et al.(1996)] {aug96}
Augusteijn, T., van der Hooft, F., de Jong, J. A., \& van Paradijs, J. 1996, \aap, 311, 889
\bibitem [Augusteijn et al.(1993)] {aug93}
Augusteijn, T., van Kerkwijk, M. H., \& van Paradijs, J. 1993, \aap, 267, L55
\bibitem [Baraffe et al.(1997)] {bar97}
Baraffe, I., Chabrier, G., Allard, F., \& Hauschildt, P. 1997, \aap, 327, 1054
\bibitem [Baraffe et al.(1998)] {bar98}
Baraffe, I., Chabrier, G., Allard, F., \& Hauschildt, P. 1998, \aap, 337, 403
\bibitem [Baraffe et al.(2003)] {bar03}
Baraffe et al. 2003, \aap, 402, 701
\bibitem [B\'ejar et al.(2001)] {bej01}
B\'ejar, V.J.S., et al. 2001, \apj, 556, 830
\bibitem [Burgasser et al.(1999)] {bur99}
Burgasser, A.J., et al. 1999, \apj, 522, L65
\bibitem [Bouvier et al.(1998)] {bou98}
Bouvier, J., et al. 1998, \aap, 336, 490
\bibitem [Bouvier et al.(2001)] {bou01}
Bouvier, J., et al. 2001, \aap, 375, 989
\bibitem [Burgasser et al.(2002)] {bur02}
Burgasser, A.J., et al. 2002, \apj, 564, 421
\bibitem [Burrows et al.(1997)] {bur97}
Burrows, A., Marley, M., Hubbard, W.B., Lunine, J.I., Guillot, T., Saumon, D., Freedman, R., Sudarsky, D., \& Sharp, C., 1997, \apj, 491, 856
\bibitem [Chabrier(2002)] {cha02}
Chabrier, G. 2002, \apj, 567, 304
\bibitem [Chabrier \& Baraffe(1997)] {cb97}
Chabrier, G. \& Baraffe, I. 1997, \aap, 327, 1039
\bibitem [Chabrier \& Baraffe(2000)] {cb00}
Chabrier, G. \& Baraffe, I. 2000, \araa, 38, 337 
\bibitem [Chabrier et al.(2000)] {cbet00}
Chabrier, G., Baraffe, I., Allard, F., \& Hauschildt, P.H. 2000, \apj, 542, 464
\bibitem [Close et al.(2002)] {clo02}
Close, L.M., Siegler, N., Potter, D., Brandner, W. \& Liebert, J. 2002, \apj, 567, L53
\bibitem [Dahn et al.(2002)] {dah02}
Dahn et al. 2002, \aj, 124, 1170
\bibitem [deKool(1992)] {dek92}
deKool, M. 1992, \aap, 261, 188
\bibitem [Delfosse et al.(1999)] {del99}
Delfosse, X., et al. 1999 \aaps, 135, 41
\bibitem [di Stefano \& Rappaport(1994)] {dis94}
di Stefano, R., \& Rappaport, S. 1994, \apj, 437, 733
\bibitem [Duquennoy \& Mayor(1991)] {duq91}
Duquennoy, A., \& Mayor, M. 1991, \aap, 248, 485
\bibitem [Eggleton(2001)] {egg01}
Eggleton, P.P. 2001, in Evolution of Binary and Multiple Star Systems, ASP Conf. Ser., Vol. 229, eds. Ph. Podsiadlowski, S. Rappaport, A. King, F. D'Antona, \& L. Burder, (ASP: San Francisco), p. 157 
\bibitem [Gizis et al.(2001)] {giz01}
Gizis, J.E., Kirkpatrick, J.D., Burgasser, A., Reid, I.N., Monet, D.G., Liebert, J., \& Wilson, J.C. 2001, \apj, 551, L163
\bibitem [Goldberg et al.(2003)] {gol03}
Goldberg, D., Mazeh, T., \& Latham, D.W. 2003, \apj, 591, 397
\bibitem [Hambly et al.(1999)] {ham99}
Hambly, N.C., Hodgkin, S.T., Cossburn, M.R., \& Jameson, R.F. 1999 \mnras, 303, 835
\bibitem [Han et al.(1995)] {han95}
Han, Z., Podsiadlowski, Ph., \& Eggeleton, P.P. 1995 \mnras, 272, 900
\bibitem [Harrison et al.(2003)] {har03}
Harrison et al. 2003, \aj, 125, 2609
\bibitem [Hoard et al.(2002)] {hoa02}
Hoard, D.W., Wachter, S., Clark, L.L., \& Bowers, T.P. 2002, \apj, 565, 511
\bibitem [Howell et al.(2002)] {how02}
Howell, S.B., Mason, E., Huber, M., \& Clowes, R. 2002, \aap, 395, L47
\bibitem [Howell et al.(2001)] {how01}
Howell, S.B., Nelson, L.A., \& Rappaport, S. 2001, \apj, 550, 897
\bibitem [Howell et al.(1997)] {how97}
Howell, S.B., Rappaport, S., \& Politano, M. 1997, \mnras, 297, 929
\bibitem [Iben \& Livio(1993)] {ibe93}
Iben, I., Jr., \& Livio, M. 1993, \pasp, 105, 1373
\bibitem [Iben \& Tutukov(1985)] {ibe85}
Iben, I., Jr, \& Tutukov, A.V. 1985, \apjs, 58, 661
\bibitem [Jingyao et al.(1998)] {jin98}
Jingyao et al. 1998, Ann. Shanghai Obs. 19, 235
\bibitem [King(1988)] {kin88}
King, A.R. 1988, \qjras, 29, 1
\bibitem [King \& Schenker(2002)] {kin02}
King, A., \& Schenker, K. 2002, in The Physics of Cataclysmic Variables and Related Objects, ASP Conf. Ser., Vol. 261, ed. B.T. Gaensicke, K. Beuermann, \& K. Reinsch (ASP: San Francisco), p. 233
\bibitem [Kirkpatrick et al.(1999)] {kir99}
Kirkpatrick, J.D., et al. 1999, \apj, 519, 834
\bibitem [Kirkpatrick et al.(2000)] {kir00}
Kirkpatrick, J.D., et al. 2000, \aj, 120, 447
\bibitem [Kolb(1993)] {kol93}
Kolb, U. 1993, \aap, 271, 149
\bibitem [Kolb \& Baraffe(1999)] {kol99}
Kolb, U., \& Baraffe, I. 1999, \mnras, 309, 1034
\bibitem [Littlefair et al.(2003)] {lit03}
Littlefair, S.P., Dhillon, V.S. \& Mart\'in, E.L. 2003, \mnras, 340, 264
\bibitem [Livio(1982)] {liv82}
Livio, M. 1982, \aap, 112, 190
\bibitem [Livio \& Soker(1983)] {liv83}
Livio, M. \& Soker, N. 1983, \aap, 125, L12
\bibitem [Livio \& Soker(1984)] {liv84}
Livio, M. \& Soker, N. 1984, \mnras, 208, 783
\bibitem [Luhman(2000)] {luh00}
Luhman, K.L. 2000, \apj, 544, 1044
\bibitem [Luhman et al.(2000)] {luhet00}
Luhman, K.L., Rieke, G.H., Young, E.T., Cotera, S., Chen, H., Rieke, M.J., Schneider, G., \& Thompson, R.I. 2000, \apj, 540, 1016
\bibitem [Marcy \& Butler(2000)] {marc00}
Marcy, G.W., \& Butler, R.P. 2000, \pasp, 112, 137
\bibitem [Mart\'in et al.(1998)] {mart98}
Mart\'in, E.L., Basri, G., Gallegos, J.E., Rebolo, R., Zapatero Osorio, M.R., \& B\'ejar, V.J.S. 1998, \apj, 499, L61
\bibitem [Mart\'in et al.(2000)] {mart00}
Mart\'in, E.L., Brandner, W., Bouvier, J., Luhman, K.L., Stauffer, J., Basri, G., Zapatero Osorio, M.R., \& Barrado y Navascu\'es, D. 2000, \apj, 543, 299
\bibitem [Miller \& Scalo(1979)] {mil79}
Miller, G.E., \& Scalo, J.M. 1979, \apjs, 41, 513
\bibitem [Nelemans et al.(2001)] {nel01}
Nelemans, G., Portegies-Zwart, S.F., Verbunt, F., \& Yungelson, L.R. 2001, \aap, 368, 939
\bibitem [Olech(1997)] {ole97}
Olech, A. 1997, Acta. Astr., 47, 281
\bibitem [Paczy\'nski \& Sienkiewicz(1981)] {pac81}
Paczy\'nski, B. \& Sienkiewicz, R. 1981, \apj, 248, 27.
\bibitem [Patterson(1984)] {pat84}
Patterson, J. 1984, \apjs, 54, 443
\bibitem [Patterson(1998)] {pat98}
Patterson, J. 1998, \pasp, 110, 1132
\bibitem [Pinsonneault, Andronov \& Sills(2002)] {pin02}
Pinsonneault, M.H., Andronov, N., \& Sills, A. 2002, in The Physics of Cataclysmic Variables and Related Objects, ASP Conf. Ser., Vol. 261, ed. B.T. Gaensicke, K. Beuermann, \& K. Reinsch (ASP: San Francisco), p. 208
\bibitem [Podsiadlowski, Rappaport \& Pfahl(2002)] {pod02}
Podsiadlowski, Ph., Rappaport, S., \& Pfahl, E. D. 2002, \apj, 565, 1107
\bibitem [Politano(1988)] {pol88}
Politano, M. 1988, Ph.D thesis, University of Illinois
\bibitem [Politano(1994)] {pol94}
Politano, M. 1994, in Interacting Binary Stars, ASP Conf. Ser., Vol. 56, ed. A. W. Shafter (ASP: San Francisco), p. 430
\bibitem [Politano(1996)] {pol96}
Politano, M. 1996, \apj, 465, 338
\bibitem [Politano et al.(1998)] {pol98}
Politano, M., Howell, S.B., \& Rappaport, S., 1998, in Wild stars in the Old West: Proceedings of the 13th North American Workshop on Cataclysmic Variables and Related Objects, ASP Conf. Ser., Vol. 137, eds. S. Howell, E. Kuulkers, \& C. Woodward (ASP: San Francisco), p. 207
\bibitem [Politano \& Quandt(2003)] {pol03}
Politano, M., \& Quandt, C.R. 2003, in preparation
\bibitem [Pylyser \& Savonije(1988)] {pyl88}
Pylyser, E. \& Savonije, G.J. 1988, \aap, 191, 57
\bibitem [Rappaport, Joss \& Webbink(1982)] {rap82}
Rappaport, S., Joss, P.C., \& Webbink, R.F. 1982, \apj, 254, 616
\bibitem [Rappaport, Verbunt \& Joss(1983)] {rap83}
Rappaport, S., Verbunt, F., \& Joss, P.C. 1983, \apj, 275, 713
\bibitem [Reid et al.(2001)] {rei01}
Reid, I.N., Gizis, J.E., Kirkpatrick, J.D., \& Koerner, D.W. 2001, \aj, 121, 489
\bibitem [Reid et al.(1999)] {rei99}
Reid, I.N., et al. 1999, \apj, 521, 613
\bibitem [Ritter \& Kolb(2003)] {rit03}
Ritter, H. \& Kolb, U. 2003, \aap, 404, 301
\bibitem [Sandquist et al.(2000)] {san00}
Sandquist, E.L., Taam, R.E., \& Burkert, A. 2000, \apj, 533, 621
\bibitem [Schenker \& King(2002)] {sch02}
Schenker, K., \& King, A. 2002, in The Physics of Cataclysmic Variables and Related Objects, ASP Conf. Ser., Vol. 261, ed. B.T. Gaensicke, K. Beuermann, \& K. Reinsch (ASP: San Francisco), p. 242
\bibitem [Soker et al.(1984)] {sok84}
Soker, N., Harpaz, A., \& Livio, M. 1984, \mnras, 210, 189
\bibitem [Szkody et al.(2002)] {szk02}
Szkody, P., et al. 2002, \aj, 123, 430
\bibitem [Taam \& Sandquist(2000)] {taa00}
Taam, R.E., \& Sandquist, E.L., 2000, \araa, 38, 113
\bibitem [Terndrup et al.(2000)] {tern00}
Terndrup, D.M. et al. 2000, \aj, 119, 1303
\bibitem [Terman \& Taam(1996)] {ter96}
Terman, J.L., \& Taam, R.E. 1996, \apj, 458, 692
\bibitem [Thorstensen et al.(2002)] {tho02}
Thorstensen, J.R., Fenton, W.H., Patterson, J.O., Kemp, J., Krajci, T., \& Baraffe, I. 2002, \apj, 567, L49
\bibitem [Tutukov \& Yungleson(1979)] {tut79}
Tutukov, A.V., \& Yungleson, L.R. 1979, in Mass Loss and Evolution of O-Type Stars, ed. P.S. Conti \& W.H. de Loore (Dordrecht:  Reidel), 216
\bibitem [Uemura et al.(2001)] {uem01}
Uemura, M., Ishioka, R., Kato, T., Schmeer, P., Yamaoka, H., Starkey, D., Vanmunster, T., \& Pietz, J. 2001, IAU Circ. 7747
\bibitem [Uemura et al.(2002)] {uem02}
Uemura, M. et al. 2002, \pasj, 54, 599.
\bibitem [Ulla(1994)] {ull94}
Ulla, A. 1994, Spa. Sci. Rev., 67, 241
\bibitem [Wachter et al.(2003)] {wac03}
Wachter, S. et al. 2003, \apj, 586, 1356
\bibitem [Warner(1995)] {war95}
Warner, B. 1995, Cataclysmic Variable Stars (Cambridge: Cambridge Univ. Press)
\bibitem [Yorke, Bodenheimer \& Taam(1995)] {yor95}
Yorke, H.W., Bodenheimer, P., \& Taam, R.E. 1995, \apj, 451, 308
\bibitem [Yungelson et al.(1995)] {yun95}
Yungelson, L., Livio, M., Tutukov, A., \& Kenyon, S.J. 1995, \apj, 447, 656
\bibitem [Zapatero Osorio et al.(1997)] {zap97}
Zapatero Osorio, M.R., Mart\'in, E.L., \& Rebolo, R. 1997, \aap, 323, 105
\bibitem [Zapatero Osorio et al.(2000)] {zap00}
Zapatero Osorio, M.R., B\'ejar V.J.S., Mart\'in, E.L., Rebolo, R., Barrado y Navascu\'es, D., Bailer-Jones, C.A.L., \& Mundt, R. 2000, Sciences, 290, 103

\end{thebibliography}
\end{document}